\documentclass[paper,11pt]{JHEP}
\pdfoutput=1
\usepackage{graphicx}
\usepackage[centertags]{amsmath}
\usepackage{amsfonts}
\usepackage{amssymb} 
\usepackage{amsthm}

\newcommand{\bra}{\langle}
\newcommand{\ket}{\rangle}

\def\one{{\rm 1\kern -.9mm l}}                             %

\def\beq{\begin{equation}}
\def\eeq{\end{equation}}
\def\beqa{\begin{eqnarray}}
\def\eeqa{\end{eqnarray}}
\newcommand{\eqa}{\begin{eqnarray}}
\newcommand{\ena}{\end{eqnarray}}
\newcommand{\p}{\partial}

\newcommand{\Z}{\mathbb{Z}}

\newcommand{\cE}{\mathcal{E}}

\newcommand{\cT}{\mathcal{T}}

\newcommand\NN{{\mathbb N}}

\newcommand{\bh}{\bar{h}}

\newcommand{\bL}{\bar{L}}
\newcommand{\bep}{\bar{\epsilon}}
\newcommand{\ep}{\epsilon}

\newcommand{\bE}{\bar{\cal E}}
\newcommand{\hm}{\hat{m}}
\newcommand{\G}{{\cal G}}

\newcommand{\bn}{\bar{n}}

\newcommand{\R}{{\cal R}}

\newcommand{\ga}{{\cal C}}
\newcommand{\bga}{{\bar{\cal C}}}
\newcommand{\tba}{\text{TBA}}
\newcommand{\Lid}{\text{Li}_2}
\newcommand{\Liq}{\text{Li}_4}

\newcommand{\Log}{\text{ln}}
\newcommand{\LLQ}{\text{Li}_4}
\newcommand{\LL}{\text{Li}_2}
\newcommand{\cIR}{\text{c}_{\text{IR}}}
\newcommand{\tcIR}{\tilde{\text{c}}_{\text{IR}}}
\newcommand{\cUV}{\text{c}_{\text{UV}}}
\newcommand{\M}{{\cal M}}
\newcommand{\br}{\bar{r}}

\title{Quantisation of the effective  string with TBA}
\author{ Michele  Caselle$^\ast$, 
 Davide Fioravanti$^\diamond$, Ferdinando Gliozzi$^\ast$ and 
Roberto Tateo$^\ast$\\
$^\ast$ Dipartimento di Fisica, Universit\`a di Torino\\
and Istituto Nazionale di Fisica Nucleare - sezione di Torino, \\
Via P. Giuria 1, I-10125 Torino, Italy
\\
$^\diamond$ Istituto Nazionale di Fisica Nucleare - sezione di Bologna  \\
and Dipartimento di Fisica e Astronomia, Universit\`a di Bologna,\\
Via  Irnerio 46, I- 40127 Bologna, Italy\\
\email{caselle, gliozzi, tateo@to.infn.it, fioravanti@bo.infn.it} 
}
\abstract{In presence of a static  pair of sources, the spectrum of low-lying  states of whatever confining gauge 
theory in D space-time dimensions is described, at large source separations, by an effective string theory. 
In the far infrared  the latter flows, in the static gauge, to a two-dimensional massless free-field theory. 
It is known that the Lorentz invariance of the gauge theory fixes uniquely the first few  subleading 
corrections of this free-field limit. We point out that the first allowed correction - a 
quartic polynomial in the field derivatives - is exactly the composite 
field $T\bar{T}$, built with the chiral components, $T$ and $\bar{T}$, of the energy-momentum tensor. 
This irrelevant perturbation is quantum integrable and yields, through 
the thermodynamic Bethe Ansatz (TBA), 
 the energy levels of the string which exactly coincide with the 
Nambu-Goto spectrum. We obtain this way the results recently found by Dubovsky, Flauger and Gorbenko. 
This procedure easily generalizes to any two-dimensional  CFT. It is known that the leading deviation of the Nambu-Goto 
spectrum comes from the boundary terms of the string action. We solve the TBA equations 
on an infinite strip, identify the relevant boundary parameter and verify that it modifies the 
string spectrum as expected.    
}
\begin{document}

\section{Introduction}
\label{sec:intro}
Even though a rigorous proof of quark confinement in Yang-Mills theories is still missing, 
numerical experiments and theoretical arguments leave little doubt that this 
phenomenon is associated to the formation of a thin string-like flux tube, 
the confining string, which generates, for large quark separations, 
a linearly  rising of  confining potential.

The string-like nature of the flux tube is particularly evident in the strong 
coupling region of lattice gauge theories, where the vacuum expectation value of large Wilson 
loops is 
given by a sum over certain lattice surfaces which can be considered as the 
world-sheet of the underlying confining string. When the coupling constant decreases, 
this two-dimensional system undergoes a  roughening transition 
\cite{rough1} where the sum of these surfaces diverges and the 
colour flux tube of whatever lattice gauge theory 
undergoes a  transition towards a rough phase, which is connected to the continuum limit. 
It is widely believed that 
such a phase transition  belongs to the Kosterlitz-Thouless 
universality class \cite{Kosterlitz:1973xp}. Accordingly, the renormalisation group 
equations imply that the effective string action
$S$ describing the dynamics of the flux tube in the whole rough phase 
flows at large scales towards a  
massless free-field theory. Thus, for large enough inter-quark separations 
it is not necessary to know explicitly the specific form of the effective string action $S$, but only its infrared limit
\beq
S[X]=S_{cl}+S_0[X]+\dots,
\label{frees}
\eeq
where the classical action $S_{cl}$ describes the usual perimeter-area term,
$X$ denotes the two-dimensional bosonic 
fields $X_i(\xi_1,\xi_2)$, with  $i=1,2,\dots, D-2$,     
describing the 
transverse displacements of the string with respect the configuration 
of minimal energy, $\xi_1,\xi_2$ are the coordinates on the world-sheet
and $S_0[X]$ is the Gaussian action
\beq
S_0[X]=\frac\sigma2\int d^2\xi\left(\partial_\alpha X\cdot\partial^\alpha X
\right) ~.
\label{gauss}
\eeq
This action is written in the physical or static gauge, where the only  degrees of freedom taken into account are the physical ones i.e. the transverse displacements $X_i$.   
Even in this  infrared approximation the effective string is highly predictive, indeed it predicts the leading correction to the linear quark-anti-quark potential, known as  L\"uscher term \cite{{Luscher:1980fr},{Luscher:1980ac}}
\beq
V(R)=\sigma R-\frac{\pi(D-2)}{24R}+O(1/R^2)~.
\label{freeV}
\eeq
Accurate numerical simulations have shown the validity of that
 expectation 
\cite{{Hasenbusch:1992zz},{Caselle:1996ii},
{Juge:2002br},{Lucini:2001nv},{Luscher:2002qv},{Caselle:2004er}}.
This infrared limit also accounts for the logarithmic broadening of the 
string width as a function of the inter-quark separation 
\cite{Luscher:1980iy}. This phenomenon was first observed long time ago in the 
$\Z_2$ 3D gauge theory \cite{Caselle:1995fh} and only recently, using the very efficient L\"uscher-Weisz algorithm 
\cite{Luscher:2002qv}, has also been observed in a non-abelian Yang-Mills 
theory \cite{{Gliozzi:2010zv}, {Gliozzi:2010zt}}.

In the last few years there has been a substantial progress in lattice simulations in measuring various properties of the flux tube, in particular the 
interquark potential and the energy of the excited string states
(see e.g.
\cite{{Caselle:1994df},{Caselle:2005xy},{Caselle:2006dv},{HariDass:2006pq},{Bringoltz:2008nd},{Athenodorou:2009ms},{Athenodorou:2011rx},
{Athenodorou:2010cs},{Caselle:2010pf},
{Billo:2011fd},{Mykkanen:2012dv},{Athenodorou:2013ioa}}) which is now sensible to the first few subleading 
corrections of the free-field infrared limit of the 
effective action. The latter is the sum of all the terms respecting the symmetry of the system, which in an Euclidean space is $SO(2)\times ISO(D-2)$.
The first few terms are
\beq
S=S_{cl}+S_0[X]+\sigma\int d^2\xi\left[c_2(\partial_\alpha X\cdot\partial^\alpha X)^2+c_3(\partial_\alpha X\cdot\partial^\beta X)(\partial_\beta X\cdot
\partial^\alpha X)\right]+S_{b}+\dots~,
\label{sa}
\eeq 
where $S_b$ is the boundary action characterizing the open string. Indeed quantum 
field theories on space-time manifolds with boundaries require, in general, 
the inclusion in the action of contributions localized at 
the boundary.  If the boundary is a Polyakov line in the $\xi_0$ direction, on which we assume Dirichlet boundary conditions, the first few terms are
\beq
S_b=\int d\xi_0 \left[b_1(\partial_1 X\cdot \partial_1 X)+b_2(\partial_1\partial_0 X\cdot \partial_1\partial_0 X)
+b_3(\partial_1 X\cdot \partial_1 X)^2+
\dots\right]\,.
\label{bounda}
\eeq
Of course the addition of these terms  modifies  the spectrum of the physical states. For instance the interquark potential (\ref{freeV}) becomes, at first order in $b_1$ \cite{Luscher:2002qv},
\beq
V(R)=\sigma R-\frac{\pi(D-2)}{24R}-  b_1\frac{\pi(D-2)}{6R^2}+  O(1/R^3)~.
\eeq
In 2004 L\"uscher and Weisz \cite{Luscher:2004ib}  noted that the comparison 
of the string  partition function  on a cylinder (Polyakov correlator) 
with the sum over closed string states in a Lorentz (or rotation) invariant 
theory yields strong constraints (called open-closed string duality). In particular they showed in this way that $b_1=0$. This property was then further generalized in \cite{Aharony:2009gg}. It was subsequently recognized that an 
 essential ingredient of these constraints is the 
Lorentz invariance of the bulk space-time 
\cite{Meyer:2006qx,{aks},{Aharony:2010cx}}.
 The confining  string action
   could be regarded as the effective action obtained 
from the underlying Yang-Mills theory of the confining vacuum in presence 
of a large Wilson loop by integrating out all the massive degrees of freedom 
\cite{Aharony:2010cx}.
This integration does not spoil the original Poincar\'e invariance of the 
underlying gauge theory, however this symmetry is no longer manifest, being spontaneously broken. As expected, it is realized 
through non-linear transformations of the $X^i$'s. The effective string action (\ref{sa}) should be invariant under the infinitesimal Lorentz transformation 
in the plane $(\alpha, j)$ 
\beq
\delta X^i=-\epsilon^{\alpha j}\delta^{ij}\xi_\alpha-\epsilon^{\alpha j}X_j
\p_\alpha X^i\,.
\label{nl}
\eeq 
For instance, if we apply the transformation (\ref{nl}) to the term
$S_{b_1}$ proportional to $b_1$ in the boundary action 
$S_b=S_{b_1}+S_{b2}+\dots$ we get at once
\beq
\delta(S_{b_1})= - b_1\int\epsilon^{1 i} 
d\xi_0\,\p_1X_i+{\rm higher~ order~ terms~}\not=0 \,,
\label{b1}
\eeq
thus such a  term breaks explicitly  Lorentz invariance, 
hence $b_1=0$. In a similar way \cite{Aharony:2010cx} 
it is possible to show that $b_3=0$ .
On the contrary the $b_2$ term is compatible with Lorentz invariance provided we add an infinite sequence of
terms generated by the non-linearity of the transformation. 
The associated recursion relations can be easily solved and the final expression can be written in a closed form \cite{Billo:2012da}
\beq
S_2= b_2\int d\xi_0\left[
\frac{\partial_1\partial_0X\cdot\partial_1\partial_0X}
{1+ \partial_1X\cdot\partial^1X }+\frac{(\partial_1\partial_0X\cdot
 \partial_1X)^2}{(1+\partial_1X\cdot\partial^1X)^2}\right]\,.
\label{ltwo}
\eeq 
It is easy to construct in this way  boundary terms of higher order \cite{Billo:2012da}.
Actually this procedure  was first applied to the bulk action and it was shown that the requirement of Lorentz
invariance of the infrared  free-field limit
(\ref{frees}) generates the whole Nambu-Goto (NG) action 
\cite{{aks},{Aharony:2010cx},{Gliozzi:2011hj}}. In the latter reference this method was generalized  to the
 construction of the effective action of higher dimensional extended objects as  D-branes on which other  massless 
modes, besides the $X^i$'s, are propagating. It can also be used to construct 
the allowed bulk corrections to the Nambu-Goto action \cite{{Aharony:2011gb},{Gliozzi:2012cx},{Meineri:2013ew}}; 
further informations on the bulk corrections of the NG can be found working in a gauge where the Lorentz invariance is manifest \cite{{Aharony:2011ga},{Dubovsky:2012sh}} (see also a general discussion on this argument in \cite{Aharony:2013ipa}).
 
A formal light-cone quantisation of the Nambu-Goto action \cite{Arvis:1983fp} suggested a simple Ansatz for 
the energy spectrum of the closed string of length $R$
\beq
E_{(n,\bar{n})}(R)=\sqrt{\sigma^2R^2+4\pi\sigma\left(n+\bar{n}-
\frac{D-2}{12}\right)+
\left(\frac{2\pi(n-\bar{n})}R\right)^2}\,,
\label{closedstring}
\eeq
where the integers $n,\bar{n}$  define the total energy $2\pi n/R$  $(2\pi \bar{n}/R)$ of the left (right) moving 
massless phonons. Similarly for the open string with fixed ends, 
where $n=\bar{n}$, one has
\beq
 E_{n}(R)=\sqrt{\sigma^2R^2+2\pi\sigma\left(n-\frac{D-2}{24}\right)}\,.
 \label{openstring}
\eeq
We refer to (\ref{closedstring}) and (\ref{openstring}) as the exact Nambu-Goto spectrum even if we know that 
this Ansatz is incompatible with Lorentz invariance in $D<26$ \cite{Goddard:1973qh}, except maybe in $D=3$ \cite{Mezincescu:2011iy}. 

For large enough $R$ one can expand (\ref{closedstring}) and (\ref{openstring})
in powers of $1/\sigma R^2$. Lattice simulations of confining gauge theories in 2+1 
and 3+1 dimensions show that the ground state and the first excitations of the 
confining string have an energy spectrum very close to that of NG. 
This suggests that the 
effective action constructed along the lines illustrated above can be considered 
as small perturbations of NG action that can be evaluated using a suitable regularization  and a standard expansion in perturbative diagrams 
\cite{{Luscher:2004ib},{Aharony:2010db}}. It turns out that only 
first order calculations on the parameters $c_i$ and $b_i$ are practically 
feasible. 

Recently, a new powerful non-perturbative method has been introduced in this context \cite{Dubovsky:2012sh}. It is based on the study of the $S$ matrix describing the scattering of the quanta of the string excitations, the phonons,
in the word-sheet. Assuming a reasonably simple form for $S$ it turns out that the system is quantum integrable
\cite{Dubovsky:2012wk}, hence one can apply the method of thermodynamic Bethe Ansatz (TBA) to calculate the non-perturbative spectrum
 of the effective string.
Taking the simplest form of the $S$ matrix and some  assumptions on the way of interacting of the phonons with different flavours 
(i.e. transverse indices), it turns out the  spectrum coincides with 
the exact NG spectrum of the closed string \cite{Dubovsky:2012wk}. 
  This method was also applied to describe some apparent deviations of the spectrum of the closed string \cite{Dubovsky:2013gi}.  

In the present paper we re-derive the NG spectrum starting from the observation that the first non-Gaussian correction of the string action (\ref{sa}), once the 
 coefficients $c_i$ assume the values required by  Lorentz invariance of 
the target space, namely $c_2=\frac18$ and $c_3=-\frac12$, coincide with the
composite field $T\bar{T}$, where $T$ and $\bar{T}$ are the chiral components of the energy momentum tensor $T_{\alpha\beta}$. Thus the effective string action is, 
at this perturbative order, a two-dimensional  integrable quantum field theory formed  
by a conformal field theory  (the infrared  Gaussian limit)  perturbed by 
$T\bar{T}$. We do not need any further assumption to derive, through the TBA, the NG spectrum. 

We also show that a similar spectrum emerges from a general class of CFT's perturbed by $T\bar{T}$. The energy levels for the identity primary field and its descendents coincides
 with (\ref{closedstring}) and (\ref{openstring}) once one replaces $D-2$ with the central charge $c$. The level degeneracy differs in an intriguing way: it is know that the degeneracy of the closed string grows exponentially for large $E$  as $\sim \exp(E/T_H)$, where 
$T_H=\sqrt{3\sigma/\pi(D-2)}$ -the Hagedorn temperature- coincides with the inverse of the distance $R_c$ where the ground state of the NG 
spectrum develops a tachyon, i.e. where the argument of the square root 
in (\ref{closedstring}) vanishes. A similar relationship between the position 
of the tachyonic singularity of the ground state and the degeneracy of highly excited  
states holds for general CFT's. We shall check it in the critical Ising model where this degeneracy can be  calculated exactly.     

The interplay between quantum integrability and Lorentz invariance in the target space is an intriguing issue: the first non-Gaussian contribution 
of the string action is an integrable perturbation only if $c_2/c_3=-\frac14$ as required by Lorentz invariance, 
however this does not imply that the generated NG spectrum agrees with that of  a Lorentz-invariant string theory. 
As already mentioned, a pure NG spectrum is compatible with  Lorentz invariance in the Minkowski target 
space only for $D=26$ (and  perhaps $D=3$) and perturbative calculations show that the NG spectrum deviates from that of 
a Lorentz-invariant string already at the order $1/R^5$ in $D>3$ dimensions , and starting at the order $1/R^7$ non-universal 
terms are expected to contribute to the energy levels \cite{{Aharony:2010db},{Dubovsky:2012sh},{Aharony:2013ipa}}. These contributions could be inserted in the 
TBA approach by assuming that at shorter distances other perturbations contribute, besides $T\bar{T}$.  

Actually the leading deviation of the NG spectrum  comes from the boundary action (\ref{ltwo}). 
This is also the most significant from a phenomenological point of view, being associated with the first non-NG correction of the interquark potential.
 We have indeed \cite{{Luscher:2004ib},{Aharony:2010db}}
\beq
V(R)=\sigma R-\frac{\pi(D-2)}{24R}-\frac1{2\sigma R^3}
\left(\frac{\pi(D-2)}{24}\right)^2- b_2\frac{\pi^3(D-2)}{60R^4}+  O(1/R^5)~,
\label{pot}
\eeq  
 where the third term comes from the $c_2$ and $c_3$ terms in (\ref{sa}) 
\cite{Luscher:2004ib}. This deviation of the interquark potential has been already observed in lattice gauge theories and the $b_2$ parameter has been 
evaluated \cite{{Brandt:2010bw},{Billo:2012da}}.

In this paper we introduce the  TBA equations 
describing  these boundary effects by generalizing the approach of \cite{Dubovsky:2012sh} to
an infinite strip with Dirichlet boundary conditions.
We find that also in this  case the  equations are  solvable explicitly and the 
energy spectrum is given by (\ref{openstring}). 
It is also easy to derive the corrections of the open string NG spectrum due to the boundary constants $b_i$ and we recover in particular
eq. (\ref{pot}).

Even though the non-perturbative method we used is very powerful and the calculations of the level corrections due to these coupling constants could be
 easily pushed to any perturbative order, we do not think that this formulation of the effective string is ultraviolet  complete. 
The reason is that the whole spectrum of the theory includes an infinite set of negative energy levels: because of the square root in eq. (\ref{closedstring}), the complete energy spectrum 
is actually $\{\pm E_{(n,\bar{n})}(R)\}$. The theory can be fermionized and  one may assume that the sea of negative energy levels is completely filled, however we did not 
succeeded in finding the zero-point energy produced by this sea, because of the huge degeneracy of the string states. As far as this problem is not completely solved, one should regard this 
formulation as an effective theory.

In the next Section we provide an elementary calculation of the $T\bar{T}$ nature of the quartic term  along with some 
quantum check. 
In Section \ref{sec:S-matrix}  we describe in detail the exact $S$ matrix for a critical RG flow of the Ising model in the limit 
of the massless phonons, following the method of Aliosha 
Zamoldchikov in the study of the flux between the tricritical Ising model and the critical 
Ising model and obtain the spectrum of the $T\bar{T}$ perturbed Ising model in a closed form. 
In Section \ref{sec:degeneracy} we study the degeneracy 
of the spectrum and compare it with the degeneracy of the string. 
Then in Section \ref{sec:strip} we put the theory in an infinite strip, define a 
consistent reflection factor and, in Section 
\ref{sec:boundary},  solve the boundary TBA for the open string in the 
case $D=3$ and compare it with the perturbative calculations. 
Finally, in  Appendix \ref{sec:ADE}, we show that Nambu-Goto like spectra emerge 
from a large  class of $T\bar{T}$ perturbed CFT's and
Section \ref{sec:conclusions} contains our conclusions with a  summary of the main results.

\section{The composite perturbation $T\bar{T}$ and its expectation value}
\label{sec:TbarT}
The energy-momentum tensor of the free-field theory (\ref{gauss}) can be written as
\beq
T_{\alpha\beta}=\p_\alpha X\cdot\p_\beta X-\frac12\delta_{\alpha\beta}
\left(\p^\gamma X\cdot\p_\gamma X\right)\,.
\label{Tab}
\eeq
Note that this tensor is symmetric, traceless and conserved, as it should.
Once we put in eq. (\ref{sa}) the values of $c_2$ and $c_3$ prescribed by Lorentz invariance, we have, as anticipated in the Introduction,
\beq
S=S_{cl}+S_0[X]-\frac\sigma4\int d^2\xi\, T_{\alpha\beta}T^{\alpha \beta}+S_{b}+\dots
\label{sat}
\eeq
In two-dimensional CFT it is useful to introduce the chiral components 
$T_{zz}=\frac12(T_{11}-iT_{12})$ and $T_{\bar{z}\bar{z}}=
\frac12(T_{11}+iT_{12})$ and use the normalized quantities 
$T=-2\pi\sigma T_{zz}$, $\bar{T}=-2\pi\sigma T_{\bar{z}\bar{z}}$ in such a way the operator product expansion begins with
\beq
T(z)T(w)=\frac{D-2}2\frac1{(z-w)^4}+\dots
\eeq 
and similarly for $\bar{T}$.
Thus at the end we have
\beq
S=S_{cl}+S_0[X]-\frac1{2\pi^2\sigma}\int d^2\xi\, T \bar{T}+S_{b}+\dots
\label{SN}
\eeq

There is a consistency check based on some general properties of the expectation value of the composite field $T\bar{T}$ pointed out by Sasha Zamolodchikov
\cite{Zamolodchikov:2004ce} for any two-dimensional quantum field theory.
In the particular case of a CFT on a infinite strip we have
\beq
\bra T\bar{T}\ket=\bra T\ket\bra\bar{T}\ket\,.
\label{TT}
\eeq
Under a conformal mapping $z\to w=f(z)$,  $T$ transforms as
\beq
T(z)=\left(\frac{df}{dz}\right)^2T(w)+\frac c{12} \{f,z\}\,,
\eeq
where $\{f,z\}=-2\sqrt{f'}\frac{d^2}{dz^2}\frac1{\sqrt{f'}}$ is the  Schwarzian derivative and $c$ the central charge.

Starting from the observation that $\bra T\ket=0$ in the complex $w$ plane  and that the transformation $f(z)=\exp(\pi z/R)$ maps conformally the infinite strip into the upper half plane $\Im m (w)\ge0$, we get, as it is well known,
\beq
\bra T\ket_{\rm strip}=\bra \bar{T}\ket_{\rm strip}=
\frac c{12}\{f,z\}=-\frac{c}{24} \left(\frac\pi{R}\right)^2\,,
\label{tstrip}
\eeq 
therefore in the infinite strip limit $L\to\infty$ we should find
\beq
\int d^2\xi\, \bra T \bar{T}\ket=RL\bra T\ket^2_{\rm strip}\,.
\label{check}
\eeq
On the other hand,
the vacuum expectation value of the quartic term of the string action on a cylinder -i.e. the correlator of two
 Polyakov lines- 
has been calculated many years ago \cite{Dietz:1982uc} in the $\zeta$-function regularization and more recently \cite{Luscher:2004ib} in the
 dimensional regularization. The result is
\beq
-\frac1{2\pi^2\sigma}\int d^2\xi\, \bra T \bar{T}\ket=-\frac1{2\pi^2\sigma}
\frac{(D-2)\pi^4 L}{24^2 R^3}\left[(D-4)
E_2(\tau)^2+2E_4(\tau)\right]\,,~ \tau=i \frac{L}{2R}\,,
\eeq
where $L$ is the circumference of the cylinder. We do not need the explicit expression of the Eisenstein series $E_2$ and $E_4$ because in the infinite strip limit $L\to\infty$ they become $E_2=E_4=1$. 
In this limit we recover eq. (\ref{check}).

We conclude this Section with the following remark. The free-field action (\ref{sat}), once perturbed with $T\bar{T}$, has a new energy-momentum tensor which is no longer the
 one defined in (\ref{Tab}) as it includes  a quartic polynomial in the derivatives of $X_i$. 
Inserting this new $T\bar{T}$ perturbation generates a new energy-momentum tensor made with a polynomial of higher degree, 
and so on.  It would be interesting to see whether this kind of recursion 
generates the same sequence produced by the request of Lorentz invariance 
in the target space.

\section{The exact $S$ matrix for massless flows}
\label{sec:S-matrix}
More than twenty  years ago, Aliosha Zamolodchikov~\cite{Zamolodchikov:1991vx} has proposed an interesting  variant of  the thermodynamic 
Bethe Ansatz \cite{Zamolodchikov:1989cf}  and the exact $S$ matrix approach to   
two-dimensional quantum field theory describing  
interpolating  trajectories  among  pairs of nontrivial  CFTs. The simplest instance discussed 
in~\cite{Zamolodchikov:1991vx}, concerns the line of  second order phase transitions  connecting 
the tricritical Ising model (TIM)  --identified with the conformal minimal model  $\M_{4,5}$ perturbed by  $\phi_{13}$--
to the Ising model (IM). The latter system corresponding  to the CFT    $\M_{3,4}$ underlying 
the infrared fixed point of  the RG flow.

Massless excitations confined on a infinite line or a ring   naturally separate into right and left movers. In  
this simple example,  only one species of particles is  present. The    right-right
and left-left mover scattering  is trivial, while the left-right  scattering is described by the  amplitude        
\beq
S(p,q)=\frac{2 \sigma+ i p q }{2 \sigma- i  p q }~,
\label{zamS}
\eeq
where the real parameter  $\sigma$ sets the  scale; it plays the role of string tension in the energy spectrum. In (\ref{zamS})  $p$ is the momentum of the right mover and $-q$ the 
momentum of the left mover. 
In the  limit  $\sigma \rightarrow  \infty$,  $S(p,q) \rightarrow 1$,  right and left mover 
excitations  decouple and  the  scale invariance of the model  is fully restored at $\sigma=\infty$. 
Starting from the  $S$ matrix (\ref{zamS}),  Zamolodchikov was then  able to derive the 
thermodynamic Bethe Ansatz  equations for the vacuum energy of the theory defined on a infinite cylinder  with
 circumference $R$. 
The relevant  equations are
\beq
\ep(p)= R p -\int_{0}^{\infty} \frac{dq}{2 \pi} \phi(p,q)\,\bL(q) 
,~~~
\bep(p)= R p \ - \int_{0}^{\infty} \frac{dq}{2 \pi} \phi(p,q)\,L(q)~,
\label{TBAZ}
\eeq
where $\ep(p)$ and $\bep(p)$ are the   pseudoenergies for the right and 
the left movers, respectively,
\beq
\phi(p,q)= -i \partial_q \Log \; S(p,q)~,
\eeq
and 
\beq
L(p)=\Log(1+ e^{-\ep(p)})~,~~\bL(p)= \Log(1+ e^{-\bep(p)})~.
\eeq
The  ground state energy is 
\beq
E^{(\tba)}(R)= - \frac{\pi}{6 R} c( \sqrt{\sigma} R)= - \int_{0}^{\infty}\frac{dp}{2\pi}  (L(p)+ \bL(p))~, 
\eeq
where $c(R \sqrt{\sigma})$ is the (flowing)  effective central charge with
\beq
\cUV=c_{\text{TIM}}=c(0)= \frac{7}{10}~,~~\cIR=c_{\text{IM}}=c(\infty)=\frac12~.
\eeq
The energy levels  obtained through  the TBA method  are automatically defined with respect to the vacuum energy in infinite space,
\beq
\lim_{R \rightarrow \infty} E^{(\tba)}_0(R) = 0~.
\label{vac}
\eeq
Within a pure  two-dimensional setup, the normalization   (\ref{vac}) is perfectly acceptable  but  it differs from the  perturbative definition  about
the  ultraviolet fixed point  by a  bulk vacuum contribution   $F_0 R$  which is  not analytic in the  perturbing 
parameter~\cite{Zamolodchikov:1989cf} 
\beq
E_0^{(\tba)}(R) = - \frac{\pi \cUV}{6 R} - F_0 R + \text{regular terms} ~,~(R \simeq 0)~. 
\eeq
(For the TIM $\rightarrow$  IM  massless flow,  $F_0=2 \sigma$.)  The alternative normalisation for the energy levels 
\beq
E_n(R)=  E^{(\tba)}_n(R) + F_0 R~, 
\eeq
agrees with the definition  coming from the short distance    expansion, it seems to be a more natural choice in view of a  possible    embedding  
in an   higher dimensional space and 
it  highlights the similarity 
between the  bulk energy, exactly computable  within  the  TBA  scheme,   and the linear term
in the quark-anti-quark potential (\ref{freeV}), whose origin traces back to the classical contribution $S_ {cl}$ to  
the effective  string action (\ref{sa})
\beq
E_0(R) = F_0 R -\frac{\pi \cIR}{6R}  +\dots~,~  (\sqrt{\sigma} R \gg 1)~.
\eeq
In the far infrared  limit,  equations (\ref{TBAZ})  lead to an exact asymptotic expansion 
for $E^{(\tba)}(R)$ \cite{Zamolodchikov:1991vx,Klassen:1991ze}
\begin{align}
\label{expansion}
\begin{split}
f^{(\tba)}(t)=\frac{1}{2 \pi} R E^{(\tba)}_0(R) &= 
- \frac{1}{24} - \frac{1}{48} t - \frac{1}{48} t^2 + \left( - \frac{5}{192} + \frac{49}{400}\pi^2 \right)  t^3 \\
&+\left( - \frac{7}{192} + \frac{49}{100} \pi^2 \right) t^4 +   \left( - \frac{7}{128} + \frac{441}{320} \pi^2  - \frac{2883}{245}
 \pi^4 \right) t^5  \\
&+ \left( -\frac{11}{128} + \frac{539}{160} \pi^2  - \frac{723819}{9800} \pi^4  \right) t^6 +
\dots
\end{split}
\end{align}
where $f^{(\tba)}(t)$ is the scaling function~\cite{Zamolodchikov:1991vx}  and  $t= \pi/(12 \sigma R^2)$.
The first three terms in  (\ref{expansion}) reproduce the 
large $\sigma$   expansion   of the effective action
\beq
S_{\sigma} = S_{\text{IM}} -\frac{1}{2 \pi^2 \sigma} \int d^2 \xi \, T \bar T~,
\label{Ss}
\eeq
where $S_{\text{IM}}$ is the Ising model CFT action. Correspondingly, the scaling function 
on the cylinder  admits the perturbative expansion
\begin{align}
\label{expansion1}
\begin{split}
f^{(\text{pert})}(t) &= - \frac{\cIR}{12} + \left( \frac{\cIR}{24} \right)^2 \alpha -   
\left( \frac{\cIR}{24} \right)^3 \alpha^2  + O(\alpha^3) \\
&=- \frac{1}{24} - \frac{1}{48} t - \frac{1}{48} t^2 + O(t^3)
~,~~(\alpha=-48 t)~.
\end{split}
\end{align}
The appearance in (\ref{expansion}) of coefficients with nonzero trascendentality\footnote{i.e. with a higher power of $\pi$.} at order  
$O(t^3)$ and greater  is a clear signal of contributions from other  
irrelevant operators~\cite{Zamolodchikov:1991vx}\footnote{It is interesting to observe that this new operator contributes exactly at the same order where  the Lorentz-invariant
 effective string theory admits new 
non-NG terms \cite{{Aharony:2010db},{Dubovsky:2012sh},{Aharony:2013ipa}}}.

\begin{figure}[ht]
\centering
\includegraphics[width=10cm]{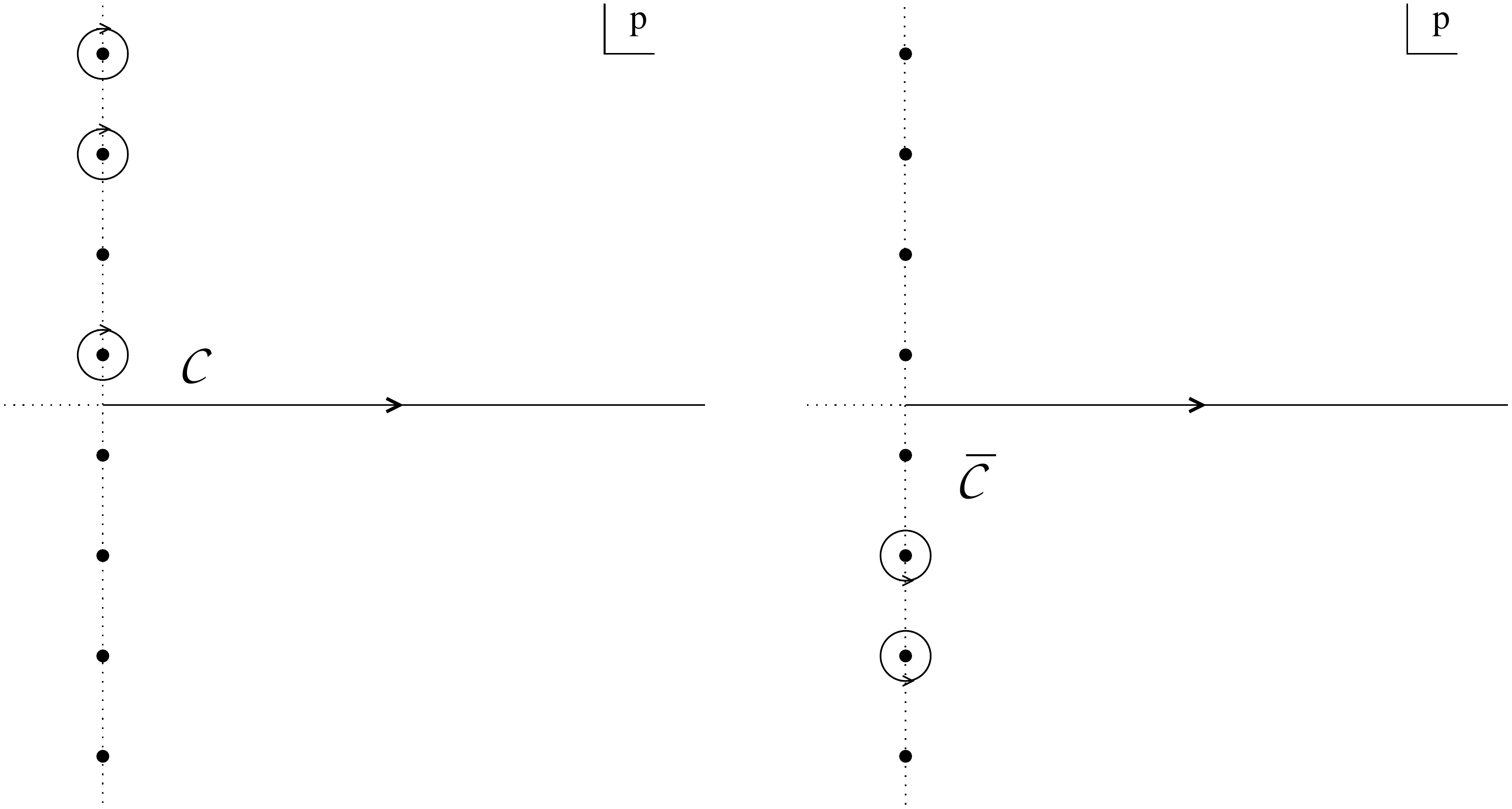}
\caption{Possible  excited state integration contours for the $\lambda_+$ sector. }
\label{cont}
\end{figure}

However, what had not been realised until  the work \cite{Dubovsky:2012sh} was that  certain   
TBA equations lead to  the full Nambu-Goto closed string spectrum. 
One of the main  achievements   of the present paper  is to generalize the results of \cite{Dubovsky:2012sh}  to other important  families of
models by  showing  that the corresponding  TBA spectra are a direct generalization of (\ref{closedstring}) and that the leading part $S_1(p,q)$ 
of the  Zamolodchikov's $S$ matrix (\ref{zamS}) at large  $\sigma$ 
\beq
S(p,q)= e^{i pq/\sigma -i (pq/\sigma)^3/12 + \dots   } = S_1(p,q) e^{-i (pq/\sigma)^3/12 + \dots}
\eeq
selects  precisely the  zero trascendentality terms in (\ref{expansion}) which  form the 
large $R$ expansion of the NG  ground state energy. Therefore, following \cite{Dubovsky:2012sh}, we replace the kernel in (\ref{TBAZ}) with 
\beq
\phi(p,q)=-i \partial_q \Log\; S_1(p,q) =  p/\sigma~,
\label{SL}
\eeq
the resulting TBA equations are
\begin{align}
\label{tbaa}
\begin{split}
\ep(p)&= R p -  \frac{p}{\sigma}  \int_{\bga} \frac{dq}{2 \pi} \bL(q)= R p +  \frac{p}{\sigma}  
\int_{\bga} \frac{dq}{2 \pi} q \partial_q\bL(q)~,~\\
\bep(p)&= R p -  \frac{p}{\sigma} \int_{\ga} \frac{dq}{2 \pi} L(q)
= R p +  \frac{p}{\sigma} \int_{\ga} \frac{dq}{2 \pi} q \partial_q L(q)~,
\end{split}
\end{align}
with 
\beq
L(q)=\Log_{\ga}(1 + \lambda_{\pm}   e^{-\ep(q)})~,~~\bL(q)= \Log_{\bga}(1 + \lambda_{\pm} e^{-\bep(q)})~.
\label{LL}
\eeq
In (\ref{LL})  $\lambda_ {+}=1$   selects the descendents of the identity and energy primary fields, 
while  $\lambda_ {-}=-1$   selects the conformal family of the  spin field~\cite{Dorey:1996re}. 
$\Log_{\ga}$ is the  continuous branch  logarithm,  $\ga$ and $\bga$ are certain integration contours running  from $q=0$ to $q=\infty$  on the real axis 
for the ground states in each subsector $\lambda_{\pm}$, but for excited states  they circle around a finite number of poles  $\{ q_i \}$ and $\{ \bar{q}_i \}$  
 of $\partial_q L(q)$ and $\partial_q \bL(q)$ (see  Figure~\ref{cont}) :   
\beq 
\ep(q_j) = i \pi (2 n_j+ (1 - \lambda_{\pm})/2)~,~~~
\bep(\bar{q}_j) = - i \pi(2  \bar{n}_j + (1 - \lambda_{\pm})/2)~,
~~(n_j, \bar{n}_j \in \NN)~.  
\eeq
Setting
\beq
\ep(p) = R p \kappa~,~~\bep(p) = R p \bar{\kappa}~,
\eeq
we find  the constraints
\beq
\kappa=  1 + \frac{4}{\pi \bar{\kappa} \sigma R^2} \LL(- \lambda_{\pm}, \bga)~,~~
\bar{\kappa}=  1 + \frac{4}{\pi \kappa \sigma R^2} 
\LL(-\lambda_{\pm}, \ga)~.
\label{kk}
\eeq
In (\ref{kk})   $\LL(z,\ga)$ denotes the  continuous branch  dilogarithm  (see e.g. \cite{Vepstas}): 
\beq
\LL(z,\ga) = -\int_{\ga} \frac{dq}{2 \pi} \Log_{\ga}(1 -z    e^{-q}) = \Lid(z) + 4 \pi^2 m - i 2 \pi n \Log(z)~,~(m,n \in \NN)   
\eeq
with $\Lid(-1)= -\frac{\pi^2}{12}$, $\Lid(1)= \frac{\pi^2}{6}$,
\begin{equation}
 \LL(-1, \ga) =\left\{
\begin{aligned}
 - &\frac{\pi^2}{6}( \cIR -24 h_{(0,0)}  - 24 n)~,~(n=0,2,3, \dots)  \\
 - &\frac{\pi^2}{6}(\cIR - 24 h_{(1,3)} - 24 n) ~,~(n=0,1,2, \dots) 
\end{aligned}\right.
\end{equation}
and
\beq
\LL(1, \ga) = \Lid(1) +  4 \pi^2 n = - \frac{\pi^2}{6}( \cIR - 24 h_{(1,2)} - 24 n)~,~(n=0,1,2, \dots) 
\eeq
where 
\beq
n=\sum_j n_j~,~\bar{n}=\sum_j \bn_j~,~~
\label{ipartition}
\eeq
and $h_{(0,0)}=0$, $h_{(1,2)}=\frac{1}{16}$, $h_{(1,3)}=\frac{1}{2}$ are the  conformal dimensions of the primary fields of  the 
minimal model $\M_{3,4}$. Therefore
\begin{align}
\begin{split}
E^{(\tba)}_{(n,\bn)}(R) &+ \sigma R = \sigma R ( \kappa+ \bar{\kappa} -1) \\
&= \sqrt{ \sigma^2 R^2 
+ \frac{\sigma}{\pi} (\LL(-\lambda_{\pm},\ga) +\LL(-\lambda_{\pm},\bga)) 
+  \left(\frac{\LL(-\lambda_{\pm}, \ga)  -\LL(-\lambda_{\pm},\bga) }{2 \pi R}\right)^2
}\\
&=\sqrt{ \sigma^2 R^2 
+ 4 \pi \sigma \left( n+\bar{n} - \frac{\tcIR}{12} \right)
+ 
\left( \frac{ 2 \pi ( n -\bar{n})}{R} \right)^2 
}~,
\label{ETBAp}
\end{split}
\end{align}
with $\tcIR= \cIR -24 h$,  ($h \in \{h_{(0,0)},  h_{(1,2)}, h_{(1,3)} \} $), 
or 
\beq
\tilde{E}^{(\tba)}_{(n,\bn)}(R)= -2 \sigma R - E^{(\tba)}_{(n,\bn)}(R)~.
\label{ETBAm}
\eeq
The result  (\ref{ETBAp}) reproduces  precisely   the zero trascendentality  coefficients  appearing in (\ref{expansion})
\begin{align}
\label{ff}
\begin{split}
f^{(\tba)}(t) &= \frac{1}{2 \pi} R E^{(\tba)}_{(0,0)}(R) =
 \frac{1}{24 t}  - \sqrt{ \frac{1}{(24 t)^2}  - \frac{\cIR}{144 t } } \\
     &= - \frac{1}{24} -  \frac{1}{48}  t -   
\frac{1}{48}  t^2 - \frac{5}{192}  t^3 - \frac{7}{192}  t^4  - \frac{7}{128}  t^5   -  \frac{11}{128}  t^6 + \dots  
\end{split}
\end{align}
Although,  as it will be discussed in greater detail  in Section~\ref{sec:degeneracy} below and similarly to 
the cases studied in~\cite{Dubovsky:2012sh},  this simple model of quantum field theory  does not possess  a standard ultraviolet fixed point, 
 it  still seems reasonable  to identify the  
bulk contribution  with the linear term  in the  short distance expansion
\beq
\partial_R  E^{(\tba)}_{(0,0)}(R)  \simeq   -F_0 + \dots = 
 -\sigma  +\dots~,~~( \text{for}~ \tcIR \ne 0)~.  
\eeq
Adding   $ F_0 R = \sigma R$, 
a nice match with   the Nambu-Goto formula (\ref{closedstring}) at  $D-2= \tcIR$  is finally obtained: 
\beq
E_{(n,\bar{n})}(R)=E^{(\tba)}_{(n,\bar{n})}(R)+ \sigma  R~.
\label{Specf}
\eeq 
Naively, we may be  tempted to discard completely the negative energy sector 
 $\tilde{E}_{(n,\bn)}(R)=-E_{(n,\bn)}(R)$, 
however,  the two branches are not completely 
disconnected as there is  a spectral singularity 
(exceptional point) at the tachyonic critical point 
\beq
R_c=\sqrt{\frac{\pi \cIR}{3\sigma}}~,
\label{itachyon}
\eeq
in the ground state $n=\bn=0$ and, for the other  levels,  singular  points at complex  values of $R$. 
The  appearance of the negative energy sector, absent   in the original work \cite{Zamolodchikov:1991vx},  is a signal of  the somehow 
pathological nature of  the  -explicitly solvable-  CFT  perturbation considered. At the level of the TBA this fact is  a  direct consequence of  
the non-localized  form of the scattering amplitude $S_1(p,q)$  used for   the kernel   (\ref{SL}). 
Still, even with a range of validity restricted to  low energy,  the appearance of the NG spectrum in the framework of  two-dimensional integrable modes  is  a
very striking  result  that may have a highly non trivial impact to   
the study of effective strings in confining gauge theory.

The results  obtained  in this  Section, although discussed  from a slightly different perspective, heavily 
rely on \cite{Dubovsky:2012sh}. 
Our model differs from those studied in  \cite{Dubovsky:2012sh} in two ways:

\begin{itemize}
 \item In \cite{Dubovsky:2012sh}  Bose  statistics was used for the derivation of  the TBA equations. The Ising model and almost  all the known integrable models 
obey instead an exclusion principle in the momentum space even when they are unmistakably associated to Bose type Lagrangians as, for instance, 
in  the quantum affine Toda field models \cite{Braden:1989bu}. A well known exception is the theory of a single (non compactified)  free Bose field
which indeed corresponds to infrared limit of the the $D=3$ case of \cite{Dubovsky:2012sh}. 
However, even  this example  does not really  represent an exception
since an alternative  Fermi type   TBA  with an additional delta-function in the kernels 
 may fully replace the original equations. The change of variable is 
\beq
\ep_B(p) \rightarrow \ep_F(p)+ \Log (1 + e^{-\ep_F(p)})~,~~
-\Log(1 - e^{- \ep_B(p)}) \rightarrow \Log (1 + e^{-\ep_F(p)})~.
\label{BoseFermi}
\eeq

\item The general case discussed  in  \cite{Dubovsky:2012sh}, consists in $D-2$  species of particles
with left-right mover  scattering  amplitude $S_{ij}(p,q)=S_1(p,q)$,  ($i,j=1,2, \dots D-2$).  Thus  microscopically  
the $D-2$ species are actually indistinguishable as  their mutual interaction is 
totally independent from their flavour $i$ and $j$, a property difficult to understand on physical grounds.

\end{itemize}
 
In the following Section we study the degeneracy of the string levels and in 
Appendix \ref{sec:ADE},  starting 
 from a  more general family  of exact scattering theories, we shall describe  
 the  generalization of  these results to more complicated conformal field theories. 
 
\section{Spectrum degeneracy}
\label{sec:degeneracy}
The degeneracy of the energy levels (\ref{Specf}) is given by the number of 
ways of writing $n$ and $\bar{n}$ in (\ref{ipartition}) i.e. the number of decompositions of $n$ and $\bar{n}$ into distinct 
integer summands without regard to the order. This is of course the degeneracy of a free fermionic system on a circle. The generating function is
\beq
\sum_{n=0}^\infty\varphi(n)q^n=\prod_{n=1}^\infty(1+q^n)=
\frac1{\prod_{n=1}^\infty(1-q^{2n-1})}\,.
\eeq
The asymptotic behaviour of the level degeneracy for large $n$ and $\bar{n}$
is known to be
\beq
\varphi(n)\varphi(\bar{n})\simeq\varphi(n)^2=
\frac1{16\sqrt{3 n^3}}e^{2\pi\sqrt{n/3}}\,.
\eeq
For large $n\simeq\bar{n}$ the energy (\ref{Specf}) is 
$E\simeq\sqrt{8\pi\sigma n}$,
so the Ising degeneracy 
$\rho_I(n)$ is
\beq
\rho_I(n)=3\left(\frac{\pi \cT_H}{3 E}\right)^3e^{E/\cT_H} = \rho_I(E) \frac{dE}{dn} \,,
\eeq
where 
\beq
\cT_H=\sqrt{\frac{3\sigma}{\pi \cIR}}\,,
\eeq
is the    Hagedorn's temperature $\cT_H=\text{sup}(T(E))$ with 
$T(E)= 1/\partial_E \Log \rho_I(E)$
and $\cIR=\frac12$. Comparison with the tachyonic 
singularity (\ref{itachyon})  at $R_c$ we obtain, as anticipated in the Introduction, $R_c\cT_H=1$. 
Notice that the degeneracy of the energy levels of the closed string 
in $D$ dimensions is asymptotically
\beq
\rho_D(n)=12(D-2)^D\left(\frac{\pi T_H}{3 E}\right)^{D+1}e^{E/T_H}\,,
\eeq
where $T_H$ differs from $\cT_H$ by the substitution $\cIR \to  D-2$. Further details on this degeneracy as well as its  thermodynamic implications 
in lattice gauge theory at finite temperature  can be found in Appendix B
of   \cite{Caselle:2011fy}.

\section{The infinite strip}
\label{sec:strip}
 Consider a single quantum  particle confined on a  segment of length $R$. The standard quantization condition for the momentum 
$p$ of the particle is  
\beq
e^{i 2 p_i R} \R_{\alpha}(p_i) \R_{\beta}(p_i)=1~,
\label{BA1}
\eeq
where $\delta_{\alpha}(p)=-i \Log \R_{\alpha}(p) $ and $\delta_{\beta}(p)=-i \Log \R_{\beta}(p) $ are the contributions to the total phase shift  
from  the reflections  on  the left and right 
boundary, respectively. 
\begin{figure}[ht]
\centering
%\vskip - 8cm 
\includegraphics[width=8cm]{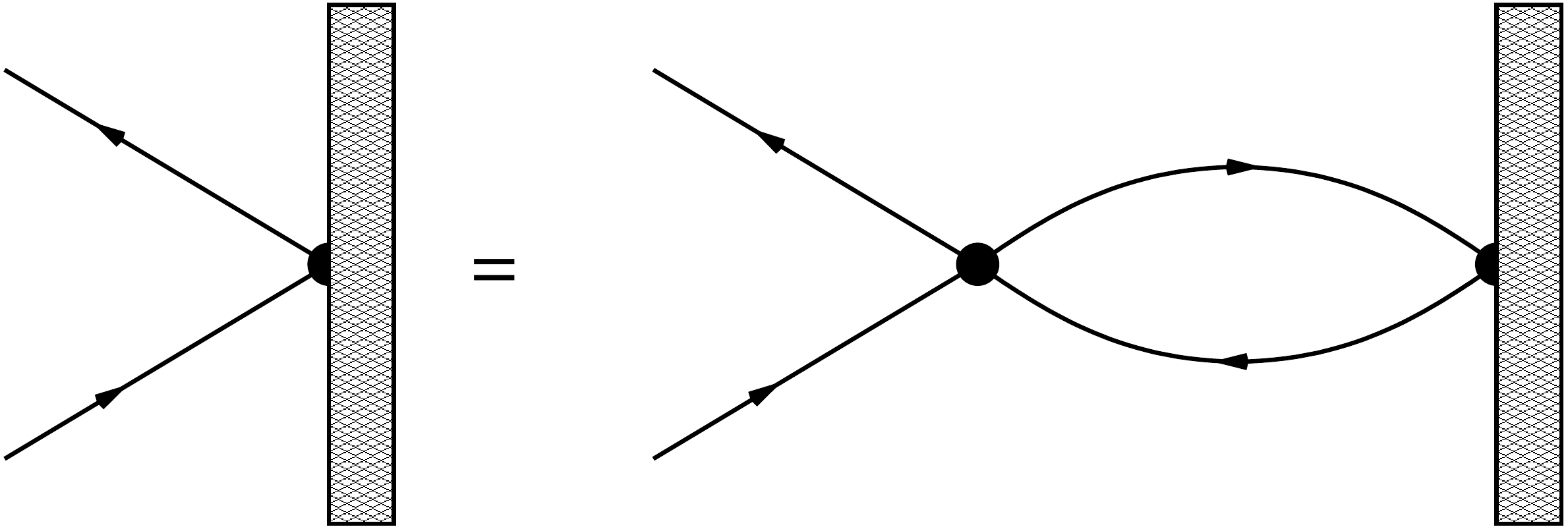}
\caption{ $\R$ matrix constraint. }
\label{RSS}
\end{figure}
Although, for an  interacting integrable two-dimensional quantum field theory confined on a space segment of length $R$ equation 
(\ref{BA1}) 
becomes   exact only in the asymptotically
large $R$ limit,  it reveals that one of the main  ingredients  for the computation of  
the spectrum is, beside the exact bulk  $S$ matrix
a consistent   reflection factor $\R(p)$.
 For massive integrable quantum field theories, the basic axiomatic  constraints linking  the boundary reflection factor $\R(p)$ to the 
two body $S$ matrix amplitude  
were discussed in  \cite{Fring:1993mp, Ghoshal:1993tm, Dorey:1998kt}. The generalization of these  results to a generic   massless perturbed conformal field theories is a nice  and partially open problem that 
certainly deserves  further attention. However a full  discussion  of this  topic   would take us too far afield, and we shall postpone it  
to the future.

For the time being, we restrict the discussion to the $D=3$ case of  \cite{Dubovsky:2012wk}, i.e. the theory 
of a single   Bose field in two dimensions   with
left-right  scattering amplitude
\beq
S_1(p,q)= e^{i p q/\sigma}~.
\label{SM}
\eeq 
Partially based on  $\text{TIM} \rightarrow  \text{IM}$ boundary  flows discussed in  \cite{Dorey:2009vg},  we identify the    relevant 
constraints to be
\beq
\R(p)\R^*(p)=1~, 
\label{U}
\eeq
which coincides with  unitary constraint for $\R(p)$,  and 
\beq
\R(p)\R(-p)=S_1(p,p)~,
\label{RS}  
\eeq  
corresponding to the equality between the   two scattering diagrams represented in Figure~\ref{RSS}.

Given the  $S$ matrix (\ref{SM}), the  minimal solution to equations (\ref{U}) and (\ref{RS}) is 
\beq
\R_0(p)=\sqrt{S_1(p,p)}=  e^{i p^2/(2 \sigma)}~,
\label{R0}
\eeq 
while  multi parameter solutions  have the general form
\beq
\R(p) =\R_0(p)   \prod_{j=1}^{\infty} \R^{(2 j - 1)}_{\delta_j}(p)~,
\label{R}
\eeq
with 
\beq
\R^{(m)}_{\delta}(p)= e^{(i p)^m \delta}~,~ (\text{m~odd})~.
\label{Rn}
\eeq
The  functions (\ref{Rn}) satisfy the simpler equation
\beq
\R^{(m)}_{\delta}(p) \R^{(m)}_{\delta}(-p)=1~. 
\label{RR1}
\eeq  
The signs of the real  parameters $\delta$'s  appearing (\ref{R}) are not constrained by  
(\ref{U}) or (\ref{RS}). 
However, as it will clear from the  results described in  Section \ref{MG} below, for left-right  symmetrical boundary  
conditions,   the  terms with the  largest  power $2j-1 \ge 3$ of $p$ and  $\delta_j \ne 0$  appearing in  (\ref{R}) should have a coefficient 
$\delta_j>0$,  to ensure a proper  convergence of the TBA integrals  and consequently also  the validity of the boundary TBA approach itself. 
 
In conclusion, the number of possible   reflection factors is  infinite. Since,  even for a simple  conformal field theory  such as the  $\M_{3,4}$ model,  the  
set of  all  possible boundary conditions corresponding to the  superposition of   Cardy's   boundary (pure) states~\cite{Cardy:1989ir} is  infinite 
dimensional, such a proliferation of free parameter  reflection factors is not  totally 
surprising.  
\section{The boundary TBA}\label{sec:boundary}
The relevant TBA equation for our simple system, obeying  Bose type  statistics\footnote{This is a conventional choice.
The equivalent Fermi statistics TBA, obtained through  the change of variable (\ref{BoseFermi}), would be fine as well.} and   defined  on a infinite strip  of size $R$ 
and  boundary conditions 
$(\alpha,\beta)$,  is~\cite{LeClair:1995uf,Dorey:1997yg}
\beq
\ep(p)= 2R p + \Lambda(p) + \frac{p}{\sigma}  \int_{\bga}  \frac{dq}{2 \pi} 
L(q)~,
\eeq
with  
\beq
L(q)=\Log_{\bga} \left(  1 - e^{-\ep(q)} \right)~,~~ \Lambda(p)=  \Log \left(\R_{\alpha}(-ip) /\R_{\beta}(ip) \right)~,
\eeq
and  vacuum and  excited state   energies  
\beq
E^{(\tba)}_n(R)=   
\int_{\bga} \frac{dp}{ 2 \pi}  L(p)~.
\eeq
\subsection{Basic boundary conditions}
Let us first consider the basic boundary conditions 
\beq
\Lambda(p)= \Log \left(\R_{0}(-ip) /\R_{0}(ip) \right)=0~, 
\eeq
the TBA equation reduces to
\beq
\ep(p)= 2R p  + \frac{p}{\sigma} \int_{\bga}  \frac{dq}{2 \pi} L(q)~.
\label{TBABB}
\eeq
The same logical steps described in Section~\ref{sec:ADE} can be repeated to find the following simple algebraic constraint 
\beq
E^{(\tba)}_n(R) = - \frac{\pi c(\bga)}{12(2R +E^{(\tba)}_n(R)/\sigma)}~,
\eeq
with $c(\bga)=1 -24 n$, ($n \in \NN$).
Considering only the positive energy solutions, we have
\beq
E^{(\tba)}_n(R) + \sigma R = \sqrt{ \sigma^2 R^2 + 2 \pi \sigma \left( n -\frac{1}{24} \right)}~.
\label{TBAff}
\eeq
Equation (\ref{TBAff}) coincides with the NG open string  spectrum (\ref{openstring}) at $D=3$.
To check the full consistency with the BA equation (\ref{BA1}), let us consider the   single particle excited state
\beq
\ep(p)= 2R p + \Log S_1(p,-i p_{j}) + \frac{p}{\sigma}  \int_{0}^{\infty}  \frac{dq}{2 \pi}
L(q)~,
\label{EBA}
\eeq
where $q=-i  p_{j}$ is the  branch point of  $L(q)$ corresponding to $\ep(-i p_{j})= -i 2 \pi n_j$, ($n_j = 1,2,\dots$).
At large $R$, setting  $\ep(i p_{j})= i 2 \pi n_j$ on the LHS of (\ref{EBA}) and 
dropping the exponentially subdominant  term on the RHS  
\beq
i 2 \pi n_j \simeq  i 2 R p_{j} + \Log S_1(i p_{j},-i p_{j}) =i 2 R p_{j} + \Log S_1(p_j,p_j)~, 
\eeq
or
\beq
e^{i 2 R p_j} S_1(p_j,p_j)=e^{i 2 R p_j} \R_0(p_j) \R_0(p_j) \simeq 1~.
\label{BA2}
\eeq
The result (\ref{BA2}) nicely fits   equation (\ref{BA1}).
\subsection{More general  boundary conditions}
\label{MG}
Let us now consider the case 
\beq
\R(p)=\R_0(p) \R_{\delta_1}^{(1)}(p)= e^{i p^2/ 2 \sigma} e^{i \delta_1 p}~,~~
\eeq
since
\beq
\Lambda(p)=  \Log \left(\R(-ip) /\R(ip) \right) =  2 p \delta_1~,
\eeq
we see that these boundary conditions simply correspond to a shift 
$R \rightarrow  R +\delta_1$, the  resulting exact spectrum is  
\beq
E^{(\tba)}_n(R,\delta_1) = E^{(\tba)}_n(R+\delta_1)~.
\eeq
Note that this case corresponds to the $b_1$ term in the boundary  action (\ref{bounda}). 
This correction was first calculated at first order in $b_1$ in \cite{Luscher:2002qv} using the 
$\zeta-$function regularization and in \cite{Luscher:2004ib} using 
dimensional regularization. In the latter reference it was also noted that 
this boundary term corresponds to a shift in $R$ and that this rule actually 
extends to the next order in $b_1$. In our approach this ``shift'' property is valid at 
any order of $\delta_1$ and the precise relation between $\delta_1$ and $b_1$ is
$ \delta_1=- 4 b_1$.

Further, we know that  Lorentz symmetry of the target space, as pointed out in (\ref{b1}), makes this term inconsistent, yet 
the TBA approach is perfectly consistent. This clearly shows that the quantum integrability does not 
imply Lorentz invariance on the target space, as anticipated in the Introduction. 

Consider now the case
\beq
\R(p)=\R_0(p) \R_{\delta_2}^{(3)}(p)= e^{i p^2/ (2 \sigma)} e^{-i p^3 \delta_2}~,~~
 \Lambda(p)=\Log \left(\R(-ip) /\R(ip) \right) = 2  p^3  \delta_2~.
\eeq
The corresponding TBA equations are
\beq
\ep(p)= 2R p +2 p^3 \delta_2  + \frac{p}{\sigma} \int_{\bga} 
\frac{dq}{2 \pi} L(q)~. 
\label{TBAw}
\eeq
The solution to (\ref{TBAw}) is of  the form
\beq
\ep(p)= 2 R \kappa p + 2  p^3 \beta~,
\eeq
with
\beq
\kappa + \frac{p^2}{R} \beta  = 1 + \frac{p^2}{R}\delta_2   + \frac{1}{2 R \sigma}
\int_{\bga} \frac{dq}{2 \pi} \Log \left( 1-e^{- 2R \kappa q - 2  q^3 \beta} \right)~. 
\eeq
Then it must be exactly $\beta=\delta_2$  and the TBA yields an exact equation for the
energy in terms of $\kappa=\kappa(\sqrt{\sigma} R, \delta_2/R^3)$
\beq
\kappa  = 1 + \frac{1}{2 R \sigma}
\int_{\bga} \frac{dq}{2 \pi} \Log \left( 1-e^{- 2R \kappa q - 2 q^3 \delta_2} \right)~, 
\eeq
which can  be easily solved  numerically  for  $R>0$ and $\delta_2>0$.
Yet,  in the literature there are interesting, large $R$, asymptotic corrections to the NG
formula  which we can match very easily. In fact, we can expand in the form
\beq
\Log \left( 1-e^{- 2R \kappa p - 2 p^3 \delta_2 } \right) =
 \Log \left( 1- e^{- 2R \kappa p} \right) + \left( \frac{2 p^3 \delta_2}{e^{2 \kappa p R} -1} \right) + \dots
\eeq
obtaining, for $\delta_2/R^3$ small,
\beq
\kappa = 1 +  \frac{1}{4 \pi \sigma R} \left(  -  \frac{\LL(1,\bga)}{2 R \kappa} +
\frac{3 \delta_2}{4 R^4 } \LLQ(1,\bga) \right) + \dots~,
\eeq
which has solution
\beq
\kappa = \frac{1}{2} \left( 1 + \sqrt{ 1- \frac{1}{2 \pi \sigma R^2 }  \LL(1,\bga)} \right) 
+ \frac{1}{4 \pi  \sigma R} \frac{3 \delta_2}{4 R^4 } \LLQ(1,\bga)+ \dots.
\eeq
The continuous branch extension   of $\Liq(1,\bga)$ is given by ~\cite{Vepstas} 
\beq
\LLQ(1,\bga )=\Liq(1)+ \frac{8}{3} \pi^4 \sum_i  (n_i)^3 = \frac{\pi^4}{90} +\frac{8}{3} \pi^4 
\sum_i  (n_i)^3~,~~ n_i \in \NN~,
\eeq
and the resulting modification to the Nambu-Goto spectrum turns out to be
\beq
E_{\{n_i \}}^{(\tba)}(R,\delta_2) = \sigma R (2 \kappa -2)=E_{n}^{(\tba)}(R)  +
\frac{\delta_2 \pi^3}{4 R^4 }  \left( \frac{1}{60} + 4 \sum_{i } (n_i)^3 \right) +\dots
\label{Edelta}
\eeq
with $n_i=0,1,\dots$, $n = \sum_i n_i$.  Equation (\ref{Edelta}) matches  the results displayed in  the Table 
on page 9 of~\cite{Aharony:2010db}, provided we set $D=3$ and identify
$ \delta_2= - 4 b_2$.

Returning  to  the positivity issue, briefly mentioned   at the end of Section \ref{sec:strip}, notice that the integral 
appearing in   equation  (\ref{TBAw}) is divergent  for $\delta_2<0$. It is therefore important
to check the sign of the latter coefficient obtained from numerical simulations.

The $n=0$ state of the spectrum  (\ref{Edelta}) was compared with
numerics in the case of the three dimensional $SU(2)$ lattice  gauge theory in \cite{Brandt:2010bw}:
$({\sigma})^{3/2} b_2 \simeq -0.015(6)(6)$. 
$b_2$  was also evaluated in \cite{Aharony:2010cx} for a  
class of holographic confining gauge theories and also in this case, with   Dirichlet's boundary conditions, the $b_2$
coefficient is  negative. Thus, in  both  cases, the TBA equation (\ref{TBAw}) 
should provide  a qualitatively good  description of the deviation of the Nambu-Goto spectrum 
(\ref{openstring}),
caused by the presence of $b_2$-perturbed boundaries, 
over a  wide  range of  $1/\sigma R^2$ about the infrared fixed  point.

Further, an high precision numerical simulation   of the three dimensional Ising gauge
model was reported  very  recently in \cite{Billo:2012da}.  The  agreement with the theoretical prediction
turned out to be very good allowing a  precise estimate of the  boundary parameter $b_2$. 
The numerical outcome,   $({\sigma})^{3/2} b_2 \simeq 0.032(2)$, leads  now  to a negative sign
for $\delta_2$. 
We interpret this result as
a clear signal of the presence of additional (infrared subleading) contributions 
associated  to extra
terms  with higher powers of $p$  in the boundary   factor $\Lambda(p)$  and/or 
in the TBA convolution kernel. We shall postpone a more complete discussion on this issue 
and a comparison between TBA and Monte Carlo results to the future.
\section{Conclusions}
\label{sec:conclusions}
In this paper we pointed out that the effective string theory describing the 
confining colour flux tube which joins a static quark-antiquark pair can be
seen as a two-dimensional CFT of central charge $D-2$  perturbed by the composite field $T\bar{T}$ made with the energy momentum tensor $T$. This perturbation 
is quantum integrable and the spectrum can be 
calculated with the TBA, as first noted in  \cite{Dubovsky:2012wk}. We 
generalized this result to a large class of conformal models. In the case 
of periodic boundary conditions the energy levels $E_{(n,\bar{n})}(R)$  are labeled by two integers
$n$ and $\bar{n}$ which depend on the monodromy of the dilogarithm in the complex plane of the momentum $p$. In a  generic ADE system  
 these energies can be parametrised in the form
$E_{(n,\bar{n})}(R)=\sigma R+\cE+\bE$, where the two quantities $\cE$ and $\bE$
obey the following consistency conditions
\beq
\cE= - \frac{\pi (\tcIR-24n)  }{12 (R + \bE/\sigma)}~,~~~
\bE= - \frac{\pi (\tcIR-24\bar{n})}{12 (R+\cE/\sigma)}\,,
\label{twosol}
\eeq
where $\tcIR$ is the effective central charge. The solution of these two
algebraic equations is exactly the NG spectrum. We then discussed the degeneracy of these states which is growing  exponentially for large $n$ and $\bar{n}$. 
Similar conclusions can be drawn for the open string case, where we wrote 
and solved the boundary TBA. We found that the reflection factor may depend on 
a set of arbitrary parameters which are associated to the coupling constants of the boundary string action. The deviation of the NG spectrum due to these 
terms can be easily calculated, at least at the first order in these 
coupling constants, and it turns out that the results coincide with those 
of the standard perturbative calculations.   
These results  give a novel perspective on the TBA, and will hopefully  lead to 
a new way to  study the interquark potential by means of    nonlinear integral equations, exact $S$ matrices and form-factors for correlation functions.
 
One of the most interesting open questions of the present approach is tied to the fact that the above equations for each pair $n,\bar{n}$ admit two 
solutions $\pm E_{(n,\bar{n})}(R)$. Thus this theory has an infinite set of negative energy levels. Even if the theory can be fermionized and we may assume 
that the sea of negative energy states is completely filled we did not succeed in evaluating the zero-point energy associated to it and its possible 
effect on the NG spectrum. One possible way out is to assume that at a given perturbative order in the $1/{\sigma R^2}$ expansion of the NG spectrum 
some other irrelevant operator starts to contribute. 
This is in particular what happens in 
the massless   flow from the tricritical Ising model to the critical one, which 
was the starting point of our analysis.

As a final remark we notice that  TBA equations  --both with  and without boundaries-- 
have emerged  in the context of ${\cal N}=4$ super Yang-Mills  
for the  study of the  quark-antiquark potential \cite{Drukker:2012de,Correa:2012hh} 
and gluon scattering amplitudes \cite{Alday:2010vh}. The latter quantities are equivalent  to 
light-like polygonal Wilson loops  and thus correspond to
the area of minimal surfaces in $AdS_5$ in the classical string theory limit. 
Although there are some  similarities between the current setup and those of 
\cite{Drukker:2012de,Correa:2012hh} and \cite{Alday:2010vh},
there are also important differences in the underlying  physics and  
the analytic properties of the corresponding TBA equations  and we are currently unable  to  identify a precise  link  
between the results of  \cite{Dubovsky:2012sh}, further developed here,  and these important  
preceding works on  $AdS_5/CFT_4$.

\medskip
\noindent{\bf Acknowledgments --}
This project was  partially supported by INFN grants IS  FI11, P14, PI11, the Italian 
MIUR-PRIN contract 2009KHZKRX-007 {\it ``Symmetries of the Universe and of the Fundamental Interactions''}, 
the UniTo-SanPaolo research  grant Nr TO-Call3-2012-0088 {\it ``Modern Applications of String Theory'' (MAST)},
the ESF Network {\it ``Holographic methods for strongly coupled systems'' (HoloGrav)} (09-RNP-092 (PESC))
 and MPNS--COST Action MP1210 {\it ``The String Theory
Universe''}.

\appendix
\section{The ADE general  case}
\label{sec:ADE}
The analysis described in  Section \ref{sec:S-matrix} can  be immediately generalised to any perturbed conformal field 
theory with  known exact $S$ matrix description, as for 
example, the massive theories represented by the diagonal reflectionless  ADE-related scattering models 
of \cite{Zamolodchikov:1989cf,Klassen:1990dx}. The TBA equations are
\beq
\ep_i(p)=  R e_i(p)  - \sum_{j=1}^{N} \int_{\ga_j} \frac{dq}{2 \pi} ~ 
\phi_{ij}(p,q)~ L_j(q)~,~~
\bep_i(p)= R e_i(p) - \sum_{j=1}^{N} \int_{\bga_{_j} } \frac{dq}{2 \pi}~
\phi_{ij}(p,q) ~\bL_j(q)~,
\label{TBAc}
\eeq
where $e_i(p)=\sqrt{p^2+m_i^2}$ is the  dispersion relation of the $i$-th particle and $N$ is the rank of  the corresponding  ADE algebra.  
The kernels  are
\beq
\phi_{ij}(p)= - i \partial_q \ln S_{ij}(p,q)~,
\eeq
where $S_{ij}(p,q)$ are  the $S$ matrix  amplitudes  of \cite{Klassen:1989ui, Braden:1989bu, Fateev:1990hy,Christe:1989ah} parametrised using  
the  momenta $p$ and $q$ of the two  particles involved in the scattering.
In the ultraviolet regime $m_iR\rightarrow 0$ the  TBA equations (\ref{TBAc}) show the decoupling of the pseudoenergies for 
right movers   from the left mover  ones
\beq
\ep_i(p)=  R \hm_i p  - \sum_{j=1}^{N} \int_{\ga_j} \frac{dq}{2 \pi} ~ 
\phi_{ij}(p,q)~ L_j(q)~,~~
\bep_i(p)= R \hm_i p - \sum_{j=1}^{N} \int_{\bga_j } \frac{dq}{2 \pi} ~
\phi_{ij}(p,q) ~\bL_j(q)~.
\label{TBAd}
\eeq
In this limit,  the energy can be found exactly~\cite{Zamolodchikov:1989cf}. The set of pure 
numbers $\hm_i\equiv m_i/m_1$   
and $i=1,\dots,N$  fix a relative scale among the particle species. They cannot be 
arbitrary numbers, but must be proportional to  the components of the 
Perron-Frobeniuns eigenvector of the corresponding Cartan matrix. 
For the vacuum  states  all this has been analysed in  \cite{Zamolodchikov:1989cf,Klassen:1990dx}, 
furthermore here we wish to take into consideration excited states as well~\cite{Bazhanov:1996aq, Dorey:1996re, Dorey:1997rb} by  
introducing complex  contours $\ga_i$ and $\bga_i$ for the continuous branch  dilogarithm and 
fugacities $\{ \lambda_i \}$ inside the statistical functions 
\beq
L_i(p)= \Log_{\ga_i} (1 + \lambda_i e^{-\ep_i(p)})~,~~
\bL_i(p)= \Log_{\bga_i} (1 + \lambda_i e^{-\bep_i(p)}) \,\, .
\eeq
The fugacities   are all equal to unity for sectors related to 
the CFT identity operator, while they may assume different values for conformal families of   other
primary fields~\cite{Martins:1991hw, Fendley:1991xn, Dorey:1996re}. In the ultraviolet limit the energy is  
\beq
E^{(\tba)}_{(n, \bn)}(R) =   \cE +\bE~,~~ \cE= - \frac{\pi}{12 R} c(\ga) ~,~~ \bE= 
- \frac{\pi}{12R} c(\bga) ~,
\eeq
where the constants 
\beq
c(\ga) =  \frac{12}{\pi}\sum_{i=1}^N     R \hm_i    \int_{\ga_i} \frac{dp}{2 \pi}  L_i(p)~,~~  
c(\bga) =  \frac{12}{\pi} \sum_{i=1}^N R \hm_i \int_{\bga_i} \frac{dp}{2 \pi} \bL_i(p)~,
\label{ch}
\eeq
can be written in terms  of the solutions to (\ref{TBAc}) and  
computed exactly using the dilogarithm trick~\cite{Zamolodchikov:1989cf, Kuniba:1992ev}. Besides, they are easily related to the conformal central 
charge $\cIR$ and the  conformal weights $(h,\bh)$ of the primary fields
\beq
c(\ga)= \tcIR- 24 n~,~~c(\bga)=\tcIR - 24 \bn~,~~ (n,\bn \in \NN)~,
\label{ch1}
\eeq 
and $\tcIR =  \cIR -24(h+\bar{h})$.
The central charge and the conformal dimensions are those for the  coset models~\cite{Goddard:1984vk}
\beq
\frac{\hat{\G}_1 \times  \hat{\G}_1}{   \hat{\G}_2}~,~~\G \in A_N, D_N, E_N~.
\eeq
Let us now come to the announced generalisation of  the analysis described in 
Section \ref{sec:S-matrix}, and  introduce
the following  variant of massless TBA equation for ADE systems
\beq
\ep_i(p)= R p \hm_i    - p  \frac{\hm_i}{\sigma} \sum_{j=1}^N  \hm_j \int_{\bga_j} 
 \frac{dq}{2 \pi}
\bL_j(q) -   \sum_{j=1}^N \int_{\ga_j} \frac{dq}{2\pi} \phi_{ij}(p,q)~ L_j(q)~,
\label{ep1}
\eeq 
\beq
\bep_i(p)= R p \hm_i   -  p \frac{\hm_i}{\sigma} \sum_{j=1}^N \hm_j  \int_{\ga_j } 
 \frac{dq}{2 \pi}
L_j(q) -  \sum_{j=1}^N \int_{\bga_j} \frac{dq}{2 \pi} \phi_{ij}(p,q)~ \bL_j(q)~.
\label{ep2}
\eeq
The elegant definition of  effective length 
\beq
r= R + \bE/\sigma~, ~~  \br=  R + \cE/\sigma~, 
\eeq 
allows to recast equations (\ref{ep1}) and  (\ref{ep2}) in the form  (\ref{TBAd}), with 
\beq
- r\cE =\frac{\pi c(\ga)}{12} =  \sum_{i=1}^N     r \hm_i    \int_{\ga_i} \frac{dp}{2 \pi}  L_i(p)~,~  
- \br \bE=\frac{\pi c(\bga)}{12} =   \sum_{i=1}^N \br \hm_i \int_{\bga_i} \frac{dp}{2 \pi} \bL_i(p)~,
\eeq
where, importantly, the $c(\ga)$ and $c(\bga)$ coincide with those
 already introduced in (\ref{ch1}) and computable  via equations (\ref{TBAd}) and (\ref{ch}).
Finally, the following self-consistent constraints  must hold
\beq
\cE= - \frac{\pi (\tcIR-24n)  }{12 (R + \bE/\sigma)}~,~~~
\bE= - \frac{\pi (\tcIR-24\bar{n})}{12 (R+\cE/\sigma)}\,.
\eeq
The latter pair of algebraic  equations  for  $\cE$ and $\bE$  can be  easily solved, giving
\beq
E_{(n,\bn)}(R)= \cE +\bE+\sigma R =\pm   \sqrt{ \sigma^2 R^2 +
4 \pi \sigma \left( n +\bn - \frac{\tcIR}{12} \right)
+\left( \frac{ 2 \pi ( n -\bar{n})}{R} \right)^2 
} \,\, . 
\label{NGfinal}
\eeq
In conclusion, we have shown that the spectrum of Nambu-Goto with $D-2=\tcIR$ emerges from a wide class of
TBA  models. 
Actually, we can   generalize this analysis  to many other interesting models, including infinite families of perturbed CFT 
theories described by non-diagonal  
$S$ matrices and we suspect that (\ref{NGfinal}) can be obtained  for  any CFT.

We end this Section  with a further  observation. It was noticed in~\cite{Zamolodchikov:2004ce} that for any  two-dimensional quantum field theory the 
expectation values of the  composite field $T \bar{T}$ 
admits an exact representation  in terms of the  expectation value of the energy-momentum tensor itself. Using the standard 
CFT convention and setting
\beq
T=-2 \pi T_{zz}~,~~ \bar{T}=-2 \pi T_{\bar{z}\bar{z}}~,~~ \Theta = 2 \pi T_{z \bar{z}}=2 \pi T_{\bar{z} z}~,
\eeq
the result of~\cite{Zamolodchikov:2004ce}, written using the double integer labeling introduced in the previous 
sections, on the cylinder is
\beq
\bra n,\bn|  T \bar{T} | n, \bn  \ket =\bra n,\bn|  T | n, \bn  \ket \bra n,\bn|  \bar{T} | n, \bn  \ket
- \bra n,\bn|  \Theta | n, \bn  \ket^2~.
\label{ttt}
\eeq
With the help of the following  relations linking the expectation values of the  energy-tensor components with the  energy 
eigenvalues    $E_{(n,\bn)}(R)$  and the total momentum of the state
$P_{(n,\bn)}(R) = 2 \pi (n-\bar{n})/R$
\begin{align}
\begin{split}
\bra n,\bn|  T_{yy}| n, \bn  \ket &= -\frac{1}{R} E_{(n,\bn)}(R)~,~~
\bra n,\bn|  T_{xx}| n, \bn  \ket = -\partial_R  E_{(n,\bn)}(R)~,~~\\ 
\bra n,\bn|  T_{xy}| n, \bn  \ket &= -\frac{i}{R} P_{(n,\bn)}(R)~,
\end{split}
\end{align}
we have~\cite{Zamolodchikov:2004ce} 
\beq
\partial_R  (E^2_{(n,\bn)}(R) - P^2_{(n,\bn)}(R)) =-\frac{ 2 R}{\pi^2} \bra n,\bn|  T \bar{T} | n, \bn  \ket~.
\label{zam}
\eeq 
Here, we would   like to remark   that inserting  espression (\ref{NGfinal}) for the energy levels in  (\ref{zam}) 
leads to 
\beq
\bra n,\bn|  T \bar{T} | n, \bn  \ket = -\pi^2 \sigma^2, 
\label{TTcost}
\eeq  
exactly and independently from the particular state  $(n,\bar{n})$  under consideration.  
Finally, using  (\ref{TTcost}) in (\ref{ttt}) gives
\beq
{\bra n,\bn|  \Theta | n, \bn  \ket} = \sqrt{
\pi^2 \sigma^2 + \bra n,\bn|  T | n, \bn  \ket \bra n,\bn|  \bar{T} | n, \bn  \ket}~.
\label{Theta}
\eeq
Since,  $\Theta=0$ corresponds to a conformal invariant theory,  
the field $\Theta$ can be identified with the CFT perturbing operator, thus 
the exact result (\ref{Theta})  should contain 
fundamental information on the further   contributions needed in ( \ref{SN}) and  (\ref{Ss}), or 
in an arbitray  
$T \bar{T}$ perturbed CFT,   to build the full action associated  to the Nambu-Goto  like spectrum (\ref{NGfinal}). 

\providecommand{\href}[2]{#2}\begingroup\raggedright\endgroup

\end{document}